\documentclass[aps,prb,twocolumn,showpacs,superscriptaddress]{revtex4}
\usepackage{amssymb}
\usepackage{amsfonts}
\usepackage{amsmath}
\usepackage{bm}
\usepackage{textcomp}
\usepackage{color}
\usepackage{longtable}
\usepackage{graphicx}
\usepackage{dcolumn}

\usepackage[a4paper=true,pagebackref=false]{hyperref}
\hypersetup{colorlinks,citecolor=blue,filecolor=blue,linkcolor=blue,urlcolor=blue}
\hypersetup{pdftitle={Magnetism of dilute (Ga,Mn)}}

\begin{document}

\title{Structural and paramagnetic properties of dilute Ga$_{1-x}$Mn$_x$N}

\author{Wiktor Stefanowicz} \affiliation{Laboratory of Magnetism,
  Bialystok University, ul. Lipowa 41, 15-424 Bialystok, Poland}
\affiliation{Institute of Physics, Polish Academy of Science,
  al.~Lotnik\'ow 32/46, PL-02-668 Warszawa, Poland}

\author{Dariusz Sztenkiel} \affiliation{Institute of Physics, Polish
  Academy of Science, al.~Lotnik\'ow 32/46, PL-02-668 Warszawa,
  Poland}

\author{Bogdan Faina} \affiliation{Institut f\"ur Halbleiter- und
  Festk\"orperphysik, Johannes Kepler University, Altenbergerstr. 69,
  A-4040 Linz, Austria}

\author{Andreas Grois} \affiliation{Institut f\"ur Halbleiter- und
  Festk\"orperphysik, Johannes Kepler University, Altenbergerstr. 69,
  A-4040 Linz, Austria}

\author{Mauro Rovezzi} \affiliation{Institut f\"ur Halbleiter- und
  Festk\"orperphysik, Johannes Kepler University, Altenbergerstr. 69,
  A-4040 Linz, Austria} \affiliation{Italian Collaborating Research
  Group, BM08 ``GILDA'', ESRF, BP 220, F-38043 Grenoble, France}

\author{Thibaut Devillers} \affiliation{Institut f\"ur Halbleiter- und
  Festk\"orperphysik, Johannes Kepler University, Altenbergerstr. 69,
  A-4040 Linz, Austria}

\author{Francesco d'Acapito} \affiliation{Consiglio Nazionale delle
  Ricerche, IOM-OGG, c/o ESRF GILDA CRG, BP 220, F-38043 Grenoble,
  France}

\author{Andrea Navarro-Quezada} \affiliation{Institut f\"ur
  Halbleiter- und Festk\"orperphysik, Johannes Kepler University,
  Altenbergerstr. 69, A-4040 Linz, Austria}

\author{Tian Li} \affiliation{Institut f\"ur Halbleiter- und
  Festk\"orperphysik, Johannes Kepler University, Altenbergerstr. 69,
  A-4040 Linz, Austria}

\author{Rafa\l \space Jakie\l a} \affiliation{Institute of Physics,
  Polish Academy of Science, al.~Lotnik\'ow 32/46, PL-02-668 Warszawa,
  Poland}

\author{Maciej Sawicki} \email{mikes@ifpan.edu.pl}
\affiliation{Institute of Physics, Polish Academy of Science,
  al.~Lotnik\'ow 32/46, PL-02-668 Warszawa, Poland}

\author{Tomasz Dietl} \email{dietl@ifpan.edu.pl}
\affiliation{Institute of Physics, Polish Academy of Science,
  al.~Lotnik\'ow 32/46, PL-02-668 Warszawa, Poland}
\affiliation{Institute of Theoretical Physics, University of Warsaw,
  PL-00-681 Warszawa, Poland}

\author{Alberta Bonanni} \email{alberta.bonanni@jku.at}
\affiliation{Institut f\"ur Halbleiter- und Festk\"orperphysik,
  Johannes Kepler University, Altenbergerstr. 69, A-4040 Linz,
  Austria}

\date{\today}

\begin{abstract}
  Systematic investigations of the structural and magnetic properties
  of single crystal Ga$_{1-x}$Mn$_x$N films grown by metal organic
  vapor phase epitaxy are presented. High resolution transmission
  electron microscopy, synchrotron x-ray diffraction, and extended
  x-ray absorption fine structure studies do not reveal any
  crystallographic phase separation and indicate that Mn occupies
  Ga-substitutional sites in the Mn concentration range up to 1\%.
  The magnetic properties as a function of temperature, magnetic field
  and its orientation with respect to the $c$-axis of the wurtzite
  structure can be quantitatively described by the paramagnetic theory
  of an ensemble of non-interacting Mn$^{3+}$ ions in the relevant
  crystal field, a conclusion consistent with the x-ray absorption
  near edge structure analysis. A negligible contribution of Mn in the
  2+ charge state points to a low concentration of residual donors in
  the studied films. Studies on modulation doped $p$-type
  Ga$_{1-x}$Mn$_x$N/(Ga,Al)N:Mg heterostructures do not reproduce the
  high temperature robust ferromagnetism reported recently for this
  system.
\end{abstract}


\maketitle
\section{Introduction}
The search for magnetic semiconductors with a Curie temperature
$T_{\mathrm{C}}$ above room temperature (RT) is currently one of the
major challenges in semiconductor
spintronics.\cite{Dietl:2000_S,Zutic:2004_RMP,Jungwirth:2006_RMP} In
single-phase samples the highest Curie temperatures reported are
$\sim$190~K for (Ga,Mn)As.\cite{Olejnik:2008_PRB,Wang:2008_APL} The
magnetic ordering in these materials is interpreted in terms of the
$p-d$ Zener model.\cite{Dietl:2000_S, Dietl:2001_PRB} This model
assumes that dilute magnetic semiconductors (DMSs) are random alloys,
where a fraction of the host cations is substitutionally replaced by
magnetic ions -- hereafter with magnetic ions we intend transition metal ions -- and the indirect magnetic coupling is provided by
delocalized or weakly localized carriers ($sp$-$d$ exchange
interactions). The authors adopted the Zener approach within the
virtual-crystal (VCA) and molecular-field (MFA) approximations with a
proper description of the valence band structure in zinc-blende and
wurtzite (wz) DMSs. The model takes into account the strong spin-orbit and the
$k \cdotp p$ couplings in the valence band as well as the influence of
strain on the band density of states. This approach describes qualitatively, and
often quantitatively the thermodynamic, micromagnetic,
transport, and spectroscopic properties of DMSs with delocalized
holes.\cite{Dietl:2004_JPCM,Jungwirth:2006_RMP}

Experimental data for Ga$_{1-x}$Mn$_x$N reveal an astonishingly wide
spectrum of magnetic properties: some groups find high temperature
ferromagnetism\cite{Reed:2001_APL,Hwang:2007_APL,Sonoda:2002_JCG} with
$T_{\mathrm{C}}$ up to 940~K,\cite{Sonoda:2002_JCG} however other
detect only a paramagnetic response and their results show that the
spin$-$spin coupling is dominated by antiferromagnetic
interactions. Generally, the origin of the ferromagnetic response in
Mn doped GaN is not clear and two basic approaches to this issue have
emerged, namely: i) methods based on the mean-field Zener
model.\cite{Dietl:2000_S} -- according to this insight, in the absence
of delocalized or weakly localized holes, no ferromagnetism is
expected for randomly distributed diluted spins.  Indeed, recent
studies of (Ga,Mn)N indicate that in samples containing up to $6\%$ of
diluted Mn, holes are strongly localized and, accordingly,
$T_{\mathrm{C}}$ below 10~K is experimentally
revealed.\cite{Sarigiannidou:2006_PRB,Edmonds:2005_APL} Higher values
of $T_{\mathrm{C}}$ could be obtained providing that efficient methods
of hole doping will be elaborated for nitride DMSs.  Surprisingly,
however, electric field controlled RT ferromagnetism has been recently
reported in Ga$_{1-x}$Mn$_x$N layers, with a Mn content as low as $x
\approx 0.25\%$.\cite{Nepal:2009_APL} These results ($T_{\mathrm{C}}
\gtrsim 300$~K) cannot be explained in the context of the $p$-$d$
Zener model, where the Curie temperature increases linearly with the
Mn concentration and for $x < 0.5\%$ $T_{\mathrm{C}}$ should not
exceed 60~K; ii) several
studies\cite{Theodoropoulou:2001_APL,Zajac:2003_JAP,Dhar:2003_PRB}
acknowledge the (likely) presence of secondary
phases -- originating from the low solubility of magnetic ions in
GaN -- as being responsible for the observation of ferromagnetism. It
has been found that the aggregation of magnetic ions leads either to
crystallographic phase separation, i.e., to the precipitation of a
magnetic compound, nanoclusters of an elemental ferromagnet, or to the
chemical phase separation into regions with respectively high and low
concentration of magnetic cations, formed without distortion of the
crystallographic structure.  It has been proposed recently that the
aggregation of magnetic ions can be controlled by varying their
valence (i.e. by tuning the Fermi level). Particularly relevant in
this context are data for (Zn,Cr)Te,\cite{Kuroda:2007_NM}
(Ga,Fe)N,\cite{Bonanni:2008_PRL} and also
(Ga,Mn)N,\cite{Kuroda:2007_NM,Reed:2005_APL,Kane:2006_JCG} where a
strict correlation between codoping, magnetic properties, and magnetic
ion distribution has been put into evidence.

There is generally a close relation between the ion arrangement
and the magnetic response of a magnetically doped semiconductor.
Specifically, depending on different preparation techniques and
parameters, coherently embedded magnetic nanocrystals [like wz-MnN
in GaN (Refs.~\onlinecite{Martinez-Criado:2005_APL} and \onlinecite{Chan:2008_PRB})] or
precipitates [like $\textit{e.g.}$ MnGa or Mn$_4$N] might in fact
give the major contribution to the total magnetic moment of the
investigated samples. In particular, randomly distributed
localized spins may account for the paramagnetic component of the
magnetization, whereas regions with a high local density of
magnetic cations are presumably responsible for ferromagnetic
features.\cite{Bonanni:2007_SST} In the case of low concentrations
of the magnetic impurity, it is often exceedingly challenging to
categorically identify the origin of the ferromagnetic signatures.

Up to very recently, in most of the reports the observation of
ferromagnetism or ferromagnetic-like behavior with apparent Curie
temperatures near or above RT, has been discussed primarily or even solely based on
magnetic hysteresis measurements. However, indirect means like superconducting quantum
interference device (SQUID)
magnetometry measurements or even the presence of the anomalous or
extraordinary Hall effect, may be not sufficient for a conclusive
statement and to verify a single-phase
system. Therefore, a careful and thorough characterization of the
systems at the nanoscale is required. This can only be achieved
through a precise correlation of the measured magnetic properties with
advanced material characterization methods, like $\textit{e.g.}$ synchrotron x-ray
diffraction (SXRD),
synchrotron based extended x-ray absorption fine structure
(EXAFS) and advanced element-specific microscopy techniques, 
suitable for the detection of a crystallographic and/or chemical phase
separation.

The present work is devoted to a comprehensive study of the
Ga$_{1-x}$Mn$_x$N ($x\leq 1\%$) fabricated by metalorganic vapor phase
epitaxy (MOVPE), which was also employed by other
authors.\cite{Nepal:2009_APL,Reed:2005_APL}. A careful on-line control
of the growth process is carried out, which is followed by an extended
investigation of the structural, optical, and magnetic properties in
order to shed new light onto the mechanisms responsible for the
magnetic response of the considered system. Particular attention is
devoted to avoid the contamination of the SQUID magnetometry signal with spurious effects
and, thus, to the reliable determination of the magnetic
properties. Experimental procedures involving SXRD, high resolution transmission electron microscopy
(HRTEM), EXAFS and x-ray absorption near-edge spectroscopy (XANES) are
employed to probe the possible presence of secondary phases,
precipitates or nanoclusters, as well as the chemical phase
separation.  Moreover, we extensively analyze the properties of single
magnetic-impurity states in the nitride host.  The understanding of
this limit is crucial when considering the most recent suggestions for
the controlled incorporation of the magnetic ions and consequently
of the magnetic response through Fermi level engineering.  By
combining the different complementary characterization techniques we establish that
randomly distributed Mn ions with a concentration $x < 1$\% generate
a paramagnetic response down to at least 2~K in Ga$_{1-x}$Mn$_x$N. In
view of our findings, the room temperature ferromagnetism observed in
this Mn concentration
range\cite{Nepal:2009_APL,Reed:2005_APL,Kane:2006_JCG,Ham:2006_ASS,Yang:2007_JCG}
has to be assigned to a non-random distribution of transition metal
impurities in GaN. We emphasize that in all reported works on (Ga,Mn)N fabricated by MOVPE
the Mn concentration was well below 5\%.

The paper is organized as follows: in the next section we give a
summary of the fabrication details, $\textit{in situ}$ monitoring of
the employed MOVPE process and an abridged overview of the
characterization techniques, together with a table listing the
principal properties and parameters characterizing the (Ga,Mn)N-based
samples considered. In Sec.~\ref{sec:structural} the results of the
structural analysis of the layers by SXRD, HRTEM, and EXAFS are
reported. These measurements prove a uniform distribution of the Mn
ions in the Ga sublattice of GaN. Section~\ref{sec:singlePhaseGaMnN}
is devoted to the determination of the Mn concentration and of the
charge and electronic state of the magnetic ions. In section
~\ref{sec:magnetic_properties} we give the experimental magnetization
characteristics of the system obtained from SQUID measurements, and
interpret the data based on the group theoretical model for Mn$^{3+}$
ions taking into account the trigonal crystal field, the Jahn-Teller
distortion and the spin-orbit coupling. Finally, conclusions and
outlook stemming from our work are summarized in
Sec.~\ref{sec:Summary}.

\section{Growth procedure}
The wz-(Ga,Mn)N epilayers here considered are fabricated by MOVPE in
an AIXTRON 200 RF horizontal reactor. All structures have been
deposited on $c$-plane sapphire substrates with TMGa (trimethylgallium),
NH$_3$, and MeCp$_2$Mn (bis-methylciclopentadienyl-manganese) as
precursors for, respectively, Ga, N and Mn, and with H$_2$ as carrier
gas. The growth process has been carried out according to a well
established procedure\cite{Bonanni:2003_JCG} consisting of: substrate
nitridation, low temperature (540$^{\circ}$C) deposition of a GaN nucleation
layer (NL), annealing of the NL under NH$_3$ until recrystallization
and the growth of a $\sim$1 $\mu$m thick device-quality GaN buffer at
1030$^{\circ}$C. On top of these structures, Mn doped GaN layers (200-700
nm) at 850$^{\circ}$C, at constant TMGa and different---over the samples
series---MeCp$_2$Mn flow-rates ranging from 25 to 490 sccm (standard
cubic centimeters per minute) have been grown. The nominal Mn content
in subsequently grown samples has been alternatively switched from low
to high and, \textit{vice versa}, to minimize long term memory effects
due to the presence of residual Mn in the reactor. During the whole
growth process the samples have been continuously rotated in order to
promote the deposition homogeneity and \textit{in situ} and on line
ellipsometry is employed for the real time control over the entire
fabrication process. The p-type superlattices have been grown
according to the optimized procedure already
reported.\cite{Simbrunner:2007_APL} Our MOVPE system is equipped with
an \textit{in situ} Isa Jobin Yvon ellipsometer that allows both
spectroscopic (variation of the optical parameters as a function of
the radiation wavelength) and kinetic (ellipsometric angles
$\textit{vs.}$ time) measurements\cite{Bonanni:2007_PRB,Peters:2000_JAP} in the energy  range 1.5 - 5.5~eV.In Table I the considered (Ga,Mn)N samples are listed together with their specific parameters.

\begingroup
\squeezetable

\begin{table}
\centering
\caption{Data related to the investigated Ga$_{1-x}$Mn$_x$N. The following values are listed: the MeCp$_2$Mn flow
rate employed to grow the Mn-doped layers, the FWHM of the
(0002) reflex from GaN determined by $\textit{ex situ}$ HRXRD, the Mn$^{3+}$
concentration as obtained from magnetization data, the total Mn content from SIMS measurements and the thickness of  each (Ga,Mn)N layer. Letters A and B denote the two different growth series}

\begin{ruledtabular}
\begin{tabular}{cccccc}


 &MeMnCp$_2$ & thickness of &  &Mn$^{3+}$ conc. & Mn conc.\\

sample & flow rate & (Ga,Mn)N & FWHM & SQUID & SIMS \\

number & [sccm] & [ nm] &[arcsec] & [$10^{20}$ cm$^{-3}$] & [$10^{20}$ cm$^{-3}$]\\

\hline
000B & 0 & 470 &  & $<$0.06 &  \\

025A & 25 & 450 & 242 & 0.28 & 0.3 \\

050A & 50 & 400 & 267 & 0.8 & 0.6 \\

100A & 100 & 400 & 243 & 0.8 & \\

100B & 100 & 520 &  & 0.27 & \\

125A & 125 & 400 & 267 & 0.6 & 0.5 \\

150A & 150 & 400 & 247 & 1.0 & 0.7\\

175A & 175 & 400 & 251 & 2.2 & \\


200B & 200 & 500 &  & 0.9 & \\

225A & 225 & 370 & 263 & 1.6 & 1.1 \\

250A & 250 & 370 & 243 & 1.4 & \\

275A & 275 & 400 & 256 & 1.6 & \\

300A & 300 & 400 & 272 & 1.4 & 1.3 \\

300B & 300 & 520 &  & 1.4 & \\

325A & 325 & 400 & 269 & 2.2 & \\

350A & 350 & 370 & 273 & 2.2 & \\

375A & 375 & 400 & 284 & 2.5 & 1.9 \\

400A & 400 & 370 & 265 & 2.6 & \\

400B & 400 & 500 &  & 2.0 & \\

475A & 475 & 700 &  & 2.7 & \\

490A & 490 & 700 &  & 3.8 & 2.4\\

490B & 490 & 470 &  & 2.7 & \\

\end{tabular}
\end{ruledtabular}

\label{tab:SampleNo}
\end{table}
\endgroup
\section{EXPERIMENTAL TECHNIQUES}
\subsection{HRTEM experimental}
HRTEM studies have been carried out on cross-sectional samples
prepared by standard mechanical polishing followed by Ar$^+$ ion
milling, under a 4$^{\circ}$ angle at 4 kV for less than 2 h. The ion
polishing has been performed in a Gatan 691 PIPS system.  The
specimens were investigated using a JEOL 2011 Fast TEM microscope
operated at 200~kV equipped with a Gatan CCD camera. The set-up is
capable of an ultimate point-to-point resolution of 0.19~nm, with the
possibility to image lattice fringes with a 0.14~nm resolution. The
chemical analysis has been accomplished with an Oxford Inca energy
dispersive x-ray spectroscopy (EDS) system.
\subsection{HRXRD and SXRD experimental}
High-resolution x-ray diffraction (HRXRD) rocking curves are routinely
acquired on each sample with a Philips XRD HR1 vertical
diffractometer with a CuK$_{\alpha}$ x-ray source working at a
wavelength of 0.15406 nm ($\sim$ 8~keV). A monochromator with a
Ge(440) crystal configuration is used to collimate the beam, that is
diffracted and collected by a Xe-gas detector. Angular ($\omega$) and
radial $\omega$/2$\theta$ scans have been collected along the growth
direction for the (002) GaN reflex, in order to gain information on
the crystal quality of the samples from the full width at half maximum
(FWHM) of the diffraction peak.

Though being aware that, if great care is exercised, also conventional
XRD may allow to detect small embedded clusters (like in the reported
case of Co in
ZnO)\cite{Venkatesan:2007_APL,Opel:2008_EPJB,Ney:2010_NJP} we
performed SXRD measurements that gave us the possibility to
additionally carry out $\textit{in situ}$ annealing experiments. The experiments
have been carried out at the beamline BM20 (Rossendorf Beam Line) of
the European Synchrotron Radiation Facility (ESRF) in Grenoble -
France. Radial coplanar scans in the $2\theta$ range from 20$^{\circ}$ to
60$^{\circ}$ were acquired at a photon energy of 10~keV. The beamline is
equipped with a double-crystal Si(111) monochromator with two
collimating/focusing mirrors (Si and Pt-coating) for rejection of
higher harmonics, allowing an acquisition energy range from 6 to 33~keV. The
measurements are performed using a heavy-duty 6-circle Huber
diffractometer, that is the system is suitable for (heavy)
user-specific environments ($\textit{e.g.}$ in our case a Be-dome for the annealing experiments
was required).

\subsection{EXAFS and XANES experimental}
The x-ray absorption fine structure (XAFS) measurements at the Mn-K
edge (6539~eV) have been performed at the GILDA Italian collaborating
research group beamline (BM08) of the ESRF in
Grenoble.\cite{D'Acapito:1998_EN} The monochromator is equipped with a
pair of Si(311) crystals and run in dynamical focusing
mode.\cite{Pascarelli:1996_JSR} Harmonics rejection is achieved
through a pair of Pd-coated mirrors with an estimated cutoff of
18~keV. Data are collected in the fluorescence mode using a 13-element
hyper pure Ge detector and normalized by measuring the incident beam
with an ion chamber filled with nitrogen gas. In order to minimize the
effects of coherent scattering from the substrate, the samples are
mounted on a dedicated sample holder for grazing-incidence
geometry;\cite{Maurizio:2009_RSI} measurements are carried out at room
temperature with an incidence angle of 1$^\circ$ and with the
polarization vector parallel to the sample surface ($E \perp c$). For
each sample the integration time for each energy point and the number
of acquired spectra are chosen in order to collect $\approx 10^6$
counts on the final averaged spectrum. Bragg diffraction peaks are
eliminated by selecting the elements of the fluorescence detector or
by manually de-glitching the affected spectra. In addition, before and
after each measurement a metallic Mn reference foil is measured in
transmission mode to check the stability of the energy scale and to
provide an accurate calibration. Considering the present optics setup
an energy resolution of $\approx 0.2$~eV is obtained at 6539~eV.

In our context, the EXAFS signal $\chi(k)$ is extracted from the
absorption raw data, $\mu(E)$, with the {\sc viper}
program\cite{Klementev:2001_JPDAP} employing a smoothing spline
algorithm and choosing the energy edge value ($E_0$) at the half
height of the energy step (Sec.~\ref{sec:xanes}). The quantitative
analysis is carried out with the {\sc ifeffit}$/${\sc artemis}
programs\cite{Newville:2001_JSR,Ravel:2005_JSR} in the frame of the
atomic model described below. Theoretical EXAFS signals are computed
with the {\sc feff8} code\cite{Ankudinov:1998_PRB} using muffin tin
potentials and the Hedin-Lunqvist approximation for their
energy-dependent part. In order to reduce the correlations between
variables, the minimum set of free fitting parameters used in the
analysis is: $\Delta$E$_0$ (correction to the energy edge), S$_0^2$
(amplitude reduction factor), $\Delta$R$_0$, $\Delta$R$_1$ (lattice
expansion factors, respectively, for the first Mn-N coordination shell
distances and all other upper distances) and $\sigma^2_i$
Debye-Waller factor for the $i^{\rm th}$ coordination shell around
the absorber plus a correlated Debye model\cite{Poiarkova:1999_PRB}
for multiple scattering paths with a fitted Debye temperature of
470(50)~K.\cite{Passler:2007_JAP}
\begin{figure}[t]
  \includegraphics[width=8.8cm]{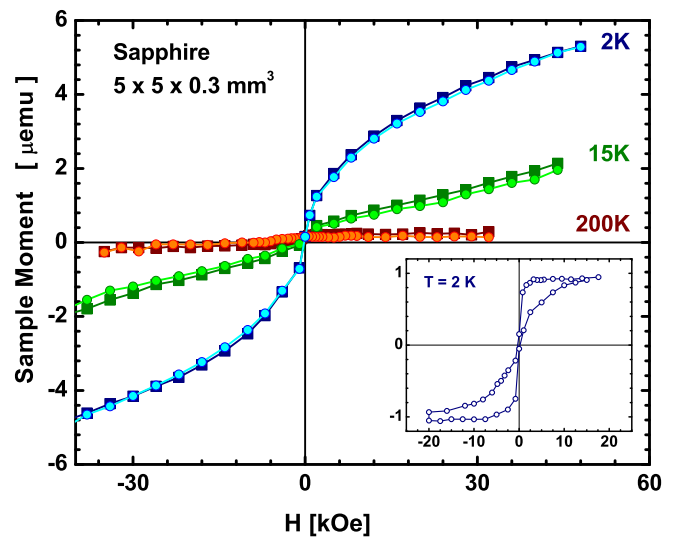}
  \caption{(Color online) Magnetic response at 2, 15, and 200 K of a
    typical ($5\times 5\times 0.3$)~mm$^3$ sapphire substrate measured
    for both in-plane (darker shade squares) and out-of-plane (lighter
    shade circles) configurations after the application of a correction linear in
    the magnetic field and proportional to the magnetic
    susceptibility of sapphire at 200~K. The magnetic moment obtained
    in this way at 2 K reaches a value that would give a magnetization
    of $\sim1$ emu/cm$^3$ for (a typical) 200~nm thick layer. Inset:
    $m(H)$ at 2~K for the same sapphire sample, but without
    correction. The axes labels are the same as in the main panel.
    This ferromagnetic-like signal is isotropic, decreases with
    temperature and vanishes above 15~K.} \label{fig:Szafiry}
\end{figure}
\subsection{SQUID experimental}
\label{sec:SQex}
The magnetic properties have been investigated in a Quantum Design
MPMS XL 5 SQUID magnetometer between 1.85 and 400~K and up to 5~T.
For magnetic studies the samples are typically cut into ($5\times 5$)
mm$^{2}$ pieces, and both in- and out-of-plane orientations are
probed. The (Ga,Mn)N layers are grown on 330 $\mu$m thick sapphire
substrates, so that the TM-doped overlayers constitute only a tiny
fraction of the volume investigated, and due to the substantial
magnetic dilution their magnetic moment is very small to small when
compared to the diamagnetic signal of the substrate.  Therefore, a
simple subtraction of a diamagnetic component originating from the
sapphire substrate and linear with the field only exposes the
resulting data to various artifacts related to the SQUID system and to
arrangement of the measurements, as already discussed in
Refs.~\onlinecite{Bonanni:2007_PRB}, \onlinecite{Salzer:2007_JMMM}, and \onlinecite{Ney:2008_JMMM}. In order to circumvent this issue, the
magnetic data presented in this paper are obtained after subtracting
the magnetic response of a sapphire substrate with dimensions
equivalent to those of the investigated sample, independently measured
on the same holders and according to the same experimental
procedure. This method, in particular, eliminates a spurious magnetic
contribution that is due to the sapphire substrate and is not-linear
with the field and, moreover, depends on the temperature. As
exemplified in Fig.~\ref{fig:Szafiry} this extra $m(T,H)$ constitutes
a nontrivial and quite sizable contribution to the signal of
interest. Additionally - as shown in the inset to this figure - the
sapphire itself may convey a ferromagnetic response to the signal at
the lowest temperatures.  We have also made sure that this method is
adequate to eliminate another weak and ferromagnetic-like contribution
appearing in the data after subtracting only the compensation linear
in the field.  This fault is caused by an inaccuracy in the value of
the magnetic field as reported by the SQUID system, which assumes that
the field acting on the sample is strictly proportional to the current
sent to the superconducting coil, and disregards the magnet remanence
due to the flux pinning inside the superconducting
windings.\cite{QD:2009_manual} This remanence in our 5~T system is as
high as -15~Oe after the field has been risen to $H > +1$~T and
results in a zero field magnetic moment of +$2\times10^{-7}$~emu for
our typical sapphire substrate. Although the value is small, it
linearly scales with the mass of the substrate and it exceeds the
magnitude of the signal expected from a submicrometer thin layer of a DMS film.
\subsection{SIMS experimental}
The overall Mn concentration in the epilayers has been evaluated via
secondary-ion mass spectroscopy (SIMS). The SIMS analysis is performed
by employing Caesium ions as the primary beam with the energy set to
5.5 keV and the beam current kept at 150 nA. The raster size is 150 x
150 $\mu m^2$ and the secondary ions are collected from a central region
of 60 $\mu m$ in a diameter. The Mn concentration is derived from MnCs+
species, and the matrix signal NCs+ was taken as reference. Mn
implanted GaN is used as a calibration standard.
\section{Structural Properties}\label{sec:structural}
As already underlined, it is necessary to ascertain if the
investigated material contains any secondary phases, nanoclusters or
precipitates. In this context, it has been realized
recently\cite{Martinez-Criado:2005_APL, Jamet:2006_NM,
  Bonanni:2007_PRB,Bonanni:2008_PRL} that the limited solubility of
transition metals in semiconductors can lead to a chemical
decomposition of the alloy, $\textit{i.e.}$ the formation of regions
with the same crystal structure of the semiconductor host, but with
respectively high and low concentration of magnetic constituents.  In
this work the structural properties of the system are analyzed by
SXRD, HRTEM and EXAFS.
\subsection{HRXRD and SXRD - results}

\begin{figure}[hb]
  \includegraphics[width=8.5 cm]{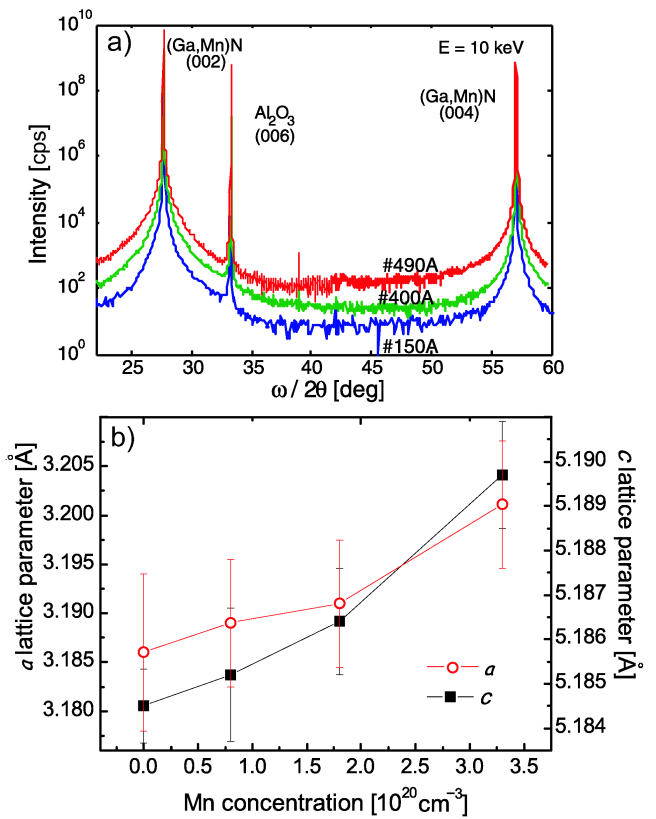}
  \caption{(Color online) a) SXRD spectra for (Ga,Mn)N samples showing
    no presence of secondary phases over a broad range of
    concentration of the magnetic ions. b) Lattice parameters
    \textit{vs.} Mn concentration. Values for an undoped GaN layer are
    added for reference}
  \label{fig:SXRD}
\end{figure}

In order to verify the homogeneity of the grown (Ga,Mn)N layers,
conventional XRD measurements have been routinely performed. From
rocking curves around the GaN (002) diffraction peak, the crystal
quality from the FWHM is verified and we obtain values in the range
240 to 290 arcsec, indicating a high degree of crystal perfection of
the layers. For the (Ga,Mn)N (002) diffraction peak we observe a shift
to lower angles with increasing Mn concentration in the acquired
$\omega/2\theta$ scans. This shift points to an increment in the
\textit{c}-lattice constant, as it has been also reported by
others.\cite{Thaler:2004_APL,Cui:2008_APL} Apart from the diffraction
peak shift, no evidence for second phases is observed in the XRD
measurements. These results have been confirmed by the SXRD
diffraction spectra reported in Fig.~\ref{fig:SXRD}(a), where no
crystallographic phase separation is detected over a broad range of Mn
concentrations.

The lattice parameters are determined by averaging the values for the
two symmetric SXRD diffractions (004) and (006) for the
\textit{c}-parameter, and one asymmetric diffraction (104) for the
\textit{a}-parameter. The variation of the lattice parameters with
increasing incorporation of Mn is presented in Fig.~\ref{fig:SXRD}b.

\begin{figure}[b]
  \includegraphics[width=8 cm]{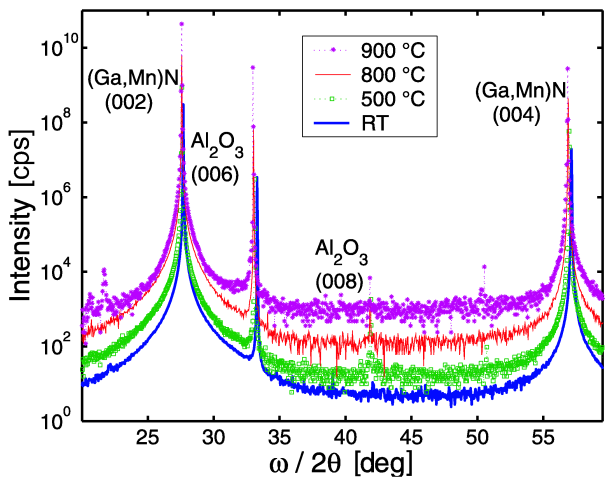}
  \caption{(Color online) \textit{In situ} SXRD spectra upon
    annealing of sample 400A at different temperatures: no formation of secondary phases is detected up to an annealing temperature of 900~$^{\circ}$C.}
  \label{fig:annealing}
\end{figure}

To obtain further information on the solubility of Mn in our (Ga,Mn)N
layers, \textit{in situ} annealing experiments have been carried out
at the ESRF BM20 beamline. Sample 400A was annealed up to 900$^{\circ}$C
in N-rich atmosphere at a pressure of 200 mbar, to compensate the
nitrogen loss during annealing. Several radial scans have been
acquired upon increasing the sample temperature in subsequent 100$^{\circ}$C
steps, and realignment was performed after reaching each
temperature. The diffraction curves upon annealing are shown in
Fig.~\ref{fig:annealing}, and no additional diffraction peaks related
to the formation of secondary phases have been detected over the whole
process.  This leads us to conclude that the considered (Ga,Mn)N grown
with Mn concentration below the solubility limit at the given
deposition conditions is stable in the dilute phase upon annealing
over a considerable thermal range.  This behavior is to be contrasted
with the one reported for dilute (Ga,Mn)As, where annealing at
elevated temperatures provokes the formation of either hexagonal or
zinc-blende MnAs nanocrystals.\cite{Moreno:2002_JAP,Tanaka:2001_JCG}
\subsection{HRTEM results}
\par
\begin{figure}[h]
  \includegraphics[width=7cm]{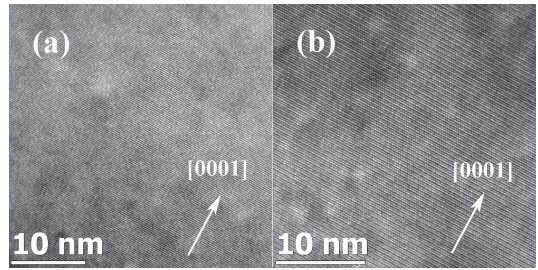}
  \caption{HRTEM images: along $[10\overline{1}0]$ (a)
    and along the $[11\overline{2}0]$ zone axis (b).}
	\label{fig:TEM}
\end{figure}

 HRTEM has been carried out on all (Ga,Mn)N layers under
consideration and independently of the Mn concentration no evidence of
crystallographic phase separation could be found. This is also
confirmed by selected area electron diffraction (SAED) patterns (not
shown) recorded on different areas of each sample, where no satellite
diffraction spot apart from wurtzite GaN are detected. In Fig.~\ref{fig:TEM}, an
example of the HRTEM images acquired along the $[10\overline{1}0]$ (a)
and $[11\overline{2}0]$ (b) zone axis, respectively is
given. Through measurements previously reported and carried out
with the same microscope, we have been able to discriminate in
(Ga,Fe)N different phases of Fe-rich nanocrystals as small as 3~nm in
diameter, and also to detect mass contrast indicating the local
aggregation of Fe-ions.\cite{Bonanni:2008_PRL,Navarro:2010_PRB} The
HRTEM images in Fig.~\ref{fig:TEM}, in contrast to the case of phase
separated (Ga,Fe)N, strongly suggest that the (Ga,Mn)N film here
studied are in the dilute state

\par
\begin{figure}[h]
  \includegraphics[width=7cm]{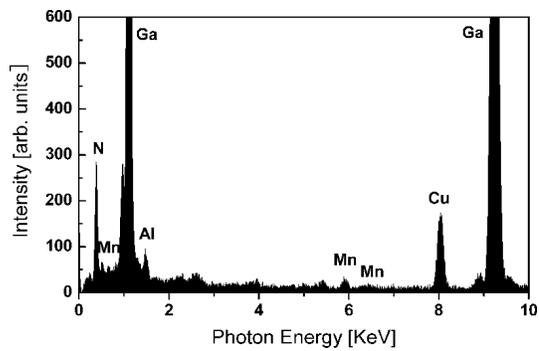}
  \caption{\label{fig:EDS} EDS spectrum of sample 300A, with the
    identification of the Mn peaks [\textit{L}$_\alpha$(0.636 keV),
    \textit{K}$_{\alpha}$(5.895 keV) and \textit{K}$_{\beta}$(6.492
    keV) ].}
\end{figure}

 The EDS spectra collected on the (Ga,Mn)N layers provide
significant signatures of the presence of Mn, as evidenced in
Fig.~\ref{fig:EDS}. The EDS detector and the software we used here
identify the Mn elements automatically, and are sensitive to Mn
concentrations as low as 0.1\% (atomic\%). The Mn concentration for
sample 300A and reported in Fig.~\ref{fig:EDS} is found to be
0.18\% (atomic\%).

\subsection{EXAFS results}\label{sec:exafs}

\begin{figure}[htbp]
  \centering
  \includegraphics[width=8.5cm]{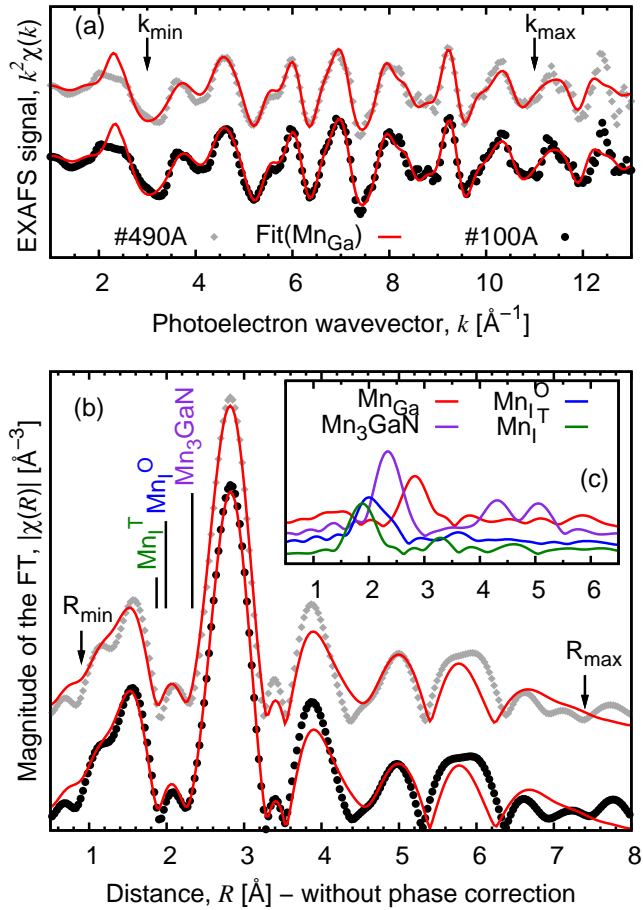}
  \caption{(Color online) $k^2$-weighted EXAFS signal, (a), for
    samples 100A (circles) and 490A (diamonds) with relative best
    fits (solid line) in the region [R$_{\rm min}$-R$_{\rm max}$] and,
    (b), amplitude of the Fourier transforms (FT) carried out in the
    range [k$_{\rm min}$-k$_{\rm max}$] by an Hanning window (slope
    parameter $d$k=1); the vertical lines indicate the position of the
    main peaks in the FT of the Mn$_{\rm I}^{\rm T}$, Mn$_{\rm I}^{\rm
      O}$ and Mn$_3$GaN additional structures. Their possible presence
    would be proptly detected as they fall in a region free from other
    peaks. {\sc feff8} simulations, (c), for the tested theoretical
    models as described in the text.}
  \label{fig:exafs}
\end{figure}

EXAFS (Ref. \onlinecite{Lee:1981_RMP}) is a well established tool in the study of semiconductor heterostructures and
nanostructures\cite{Boscherini:2008_book} and has proven its power as
a chemically sensitive local probe for the site identification and
valence state of Mn and Fe dopants in GaN
DMS.\cite{Soo:2001_APL,Sato:2002_JJAP,Biquard:2003_JS,Bacewicz:2003_JPCS,Rovezzi:2009_PRB} The crystallinity of the films and the optimal signal to noise ratio
of the collected spectra are demonstrated by the large number of
atomic shells visible and reproducible by the fits below 8~$\AA$ in
the Fourier-transformed spectra reported in Fig.~\ref{fig:exafs} for
the two representative samples 100A and 490A, respectively. In
addition, the homogeneous Mn incorporation along the layer thickness
is tested by measuring the Mn fluorescence yield (at a fixed energy of
6700 eV) as a function of the incidence angle (not shown). The EXAFS
response of these two samples is qualitatively equivalent, as
evidenced in Fig.~\ref{fig:exafs}, and this is confirmed by the
quantitative analysis. The best fits are obtained by employing a
substitutional model of one Mn at a Ga site (Mn$_{\rm Ga}$) in a
wurtzite GaN crystal (using the lattice parameters previously found by
SXRD). The possible presence of additional phases in the sample as
octahedral or tetrahedral interstitials (Mn$_{\rm I}^{\rm O}$,
Mn$_{\rm I}^{\rm T}$) in GaN or Mn$_3$GaN
clusters\cite{Giraud:2004_EPL} has been checked by carrying out fits
with a two phases model. The fraction of the Mn$_{\rm Ga}$ is found to
be 98(4)~\% for the pair Mn$_{\rm Ga}$-Mn$_3$GaN,
99(3)\% for the pair Mn$_{\rm Ga}$-Mn$_{\rm I}^{\rm O}$ and 97(3)\%
for the pair Mn$_{\rm Ga}$-Mn$_{\rm I}^{\rm T}$, respectively. With these results we
can safely rule out the occurrence of phases other than Mn$_{\rm Ga}$,
at least above 5\% level.

The local structure parameters found for the measured samples are
equivalent within the error bars (reported on the last digit within
parentheses) and averaged values are given for simplicity. The
value of the amplitude reduction factor S$_0^2$~=~0.95(5) demonstrates
the good agreement with the theoretical coordination numbers for
Mn$_{\rm Ga}$ (considering the in-plane polarization) and the
correction to the energy edge $\Delta E_0 = -7(1)$~eV supports the
XANES analysis (Sec.~\ref{sec:xanes}). With respect to the lattice
parameters previously found by SXRD, the long range distortion fits
within the error ($\Delta R_1 = 0.1(2)$~\%), while the Mn-N nearest
neighbors have a $\Delta R_0 = 2.5(5)$~\% (expansion to 1.99(1)~\AA),
in line with previously reported experimental
results\cite{Sato:2002_JJAP,Bacewicz:2003_JPCS,Biquard:2003_JS} and
recent $\textit{ab initio}$ calculations.\cite{Stroppa:2009_PRB} Finally, all
the evaluated $\sigma^2_i$ attest around the average value of $8(2)
\cdot 10^{-3}$~\AA$^{-2}$, confirming the high crystallinity of the
layers.
\section{Properties of homogeneous single-phase $\mathrm{\textbf{(Ga,Mn)N}}$} \label{sec:singlePhaseGaMnN}

Thus, SXRD, HRTEM and XAFS experiments have confirmed the wurtzite
structure of the samples, the absence of secondary phases, and the
location of Mn in the Ga sublattice of the wurtzite GaN
crystal. Furthermore, the samples have been investigated to determine
the actual Mn concentration and the charge state of the magnetic ions.
\subsection{Determination of the Mn concentration}
\label{sec:x-Mn}

\begin{figure}[htbp]
  \centering
  \includegraphics[width=8cm]{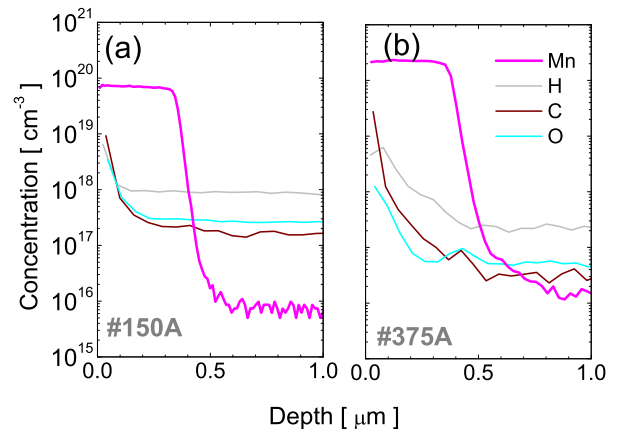}
  \caption{(Color online) SIMS depth profiles of Mn, C, O and H for
    the samples: a) 150A and b) 375A.}
  \label{fig:SIMS_thin_films}
\end{figure}

The depth profiling capabilities of SIMS provide not only an accurate
analysis of the (Ga,Mn)N layers composition, but allow also to monitor
the changes in composition along the sample depth. The SIMS depth
profiles reported in Figs.~\ref{fig:SIMS_thin_films}(a) and (b) give
evidence that the distribution of the Mn concentration  $n_{\mathrm{Mn}}$ in the
investigated films is essentially uniform over the doped layers,
independent of the magnetic ions content as well as that the interface
between the (Ga,Mn)N overlayer and the GaN buffer layer is sharp. This
is confirmed by EDS studies, which with the sensitivity around 0.1\%
at. do not provide any evidence for Mn diffusion into the buffer. The
determined total Mn concentration increases with increasing MeCp$_2$Mn
flow rate and the corresponding $n_{\mathrm{Mn}}$ values for the
considered samples can be found in Table~\ref{tab:SampleNo}.

\subsection{Energy levels introduced by Mn impurities}

The character of the paramagnetic response of DMS depends crucially on
the magnetic ion configuration. In III-V semiconductors, Mn in the
impurity limit substitutes the cation site giving three electrons to
the crystal bond. Depending on the compensation ratio, Mn can exist in
three different charge states and electron configurations, namely: i)
ionized acceptor Mn$^{2+}$, with five electrons localized in the Mn
$d$ shell. The electronic configuration of Mn$^{2+}$ is $d^{5}$, and
the ground level of the ion at zero magnetic field is a degenerate
multiplet with vanishing orbital momentum ($L=0$, $S=5/2$). The
magnetic moment of the ion results solely from the spin, and its
magnetic contribution can be described by a standard Brillouin
function for any orientation of the magnetic field. The neutral
configuration of Mn$^{3+}$ ($S=2$, $L=2$) can be realized in two ways:
ii) by substitutional manganese $d^{4}$ with four electrons tightly
bound in the Mn $d$ shell; iii) Mn$^{2+}$ + hole ($d^{5}$ + hole) with
five electrons in the Mn $d$ shell and a bound hole localized on
neighboring anions.
\subsection{XANES results}\label{sec:xanes}

The XANES spectra allow to determine the redox-state of the probed
species and give information on the structure of the surroundings of
the absorbing atom.\cite{Yamamoto:2008_XRS} Basically, the near edge
region resembles the density of those empty states, that are
accessible via optical transitions from the Mn $1s$ shell.

The goal of our XANES analysis is to determine the valence state of Mn
and to confirm the Mn$_{\rm Ga}$ incorporation, in comparison to the
findings and analysis carried out previously for molecular beam
epitaxy (MBE)-grown (Ga,Mn)N, and interpreted in terms of Mn$^{3+}$
(Refs.~\onlinecite{Sarigiannidou:2006_PRB} and \onlinecite{Titov:2005_PRB}) or
Mn$^{2+}$ (Ref.~\onlinecite{Sancho-Juan:2009_JPCM}). In order to
assign the Mn valence state, first of all we proceed with a comparison
of the position of the absorption Mn K-edge to reference compounds,
like Mn-based oxides since we do not have available data on Mn-nitrides. This
procedure was already adopted by other
groups\cite{Biquard:2003_JS,Sancho-Juan:2009_JPCM} but its reliability
could be questionable; to clarify this point {\em ab initio}
calculations are also performed.

\begin{figure}[htbp]
  \includegraphics[width=8.5cm]{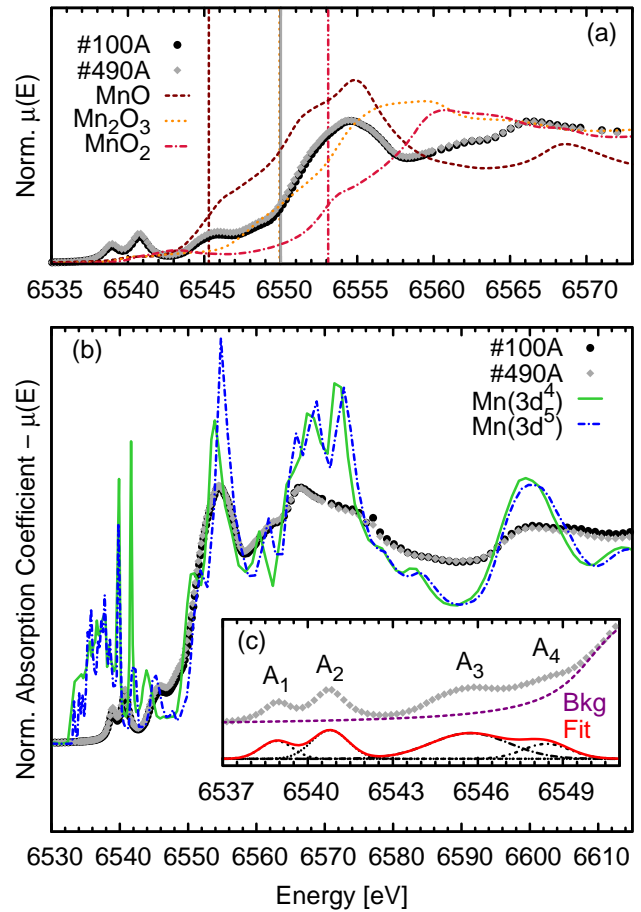}
  \caption{(Color online) Normalized XANES spectra of the samples
    100A and 490A (points) compared with: (a) the reference
    manganese oxides (MnO, Mn$_2$O$_3$, MnO$_2$) - the chosen edge
    positions are highlighted by vertical lines; (b) {\em ab initio}
    absorption spectra (without convolution) for Mn$_{\rm Ga}$ in the
    3$d^4$ and 3$d^5$ electronic configurations. The inset (c) shows
    the method used to extract the results of Table~\ref{tab:xanes},
    focusing the near-edge region for sample 490A with the baseline
    (Bkg), the relative fit and its components (A$_1$-A$_{4}$).}
  \label{fig:xanes}
\end{figure}

As shown in Fig.~\ref{fig:xanes}, the XANES spectra determined for two
samples differing in Mn concentrations (100A and 490A) are
identical, confirming a conclusion from the SQUID data on the
independence of the Mn charge state of the Mn concentration. In
Fig.~\ref{fig:xanes}$(a)$ three spectra collected in transmission mode
from commercial powders of MnO, Mn$_2$O$_3$ and MnO$_2$, with
Mn-valence states 2+, 3+, 4+, respectively, are used as reference. As
seen, with the increasing charge state, the edge moves to a higher
energy, as the accumulated positive charge shifts downwards in energy
more the $1s$ Mn shell than the valence states, in agreement with the
Haldane-Anderson rule.

Usually, the edge position is taken at the first inflection point of
the plot, but in the present case (since the oxide spectra exhibit a
broad peak that modifies the slope at the edge) a better estimate of
the edge position is obtained by considering the energy of the half
step-height of the background function. In both investigated samples
this lies at 6550.0(5)~eV. For the oxides, their half-height energies
are determined to be 6545.7(5)~eV, 6550.3(5)~eV, 6553.3(5)~eV for MnO,
Mn$_2$O$_3$ and MnO$_2$, respectively. This would strongly suggest
that we deal with Mn$^{3+}$, in line with the SQUID results
(Sec.~\ref{sec:squid}). On the other hand, taking the position of the
inflection points, the determined charge state would be 2+, as
reported in Ref.~\onlinecite{Sancho-Juan:2009_JPCM}. This demonstrates
that, in this case, relying only on the edge position to determine the
valence state is prone to error and strongly depends on the local
surrounding of the probed species.\cite{Farges:2005_PRB}

To clarify this point, we look at the pre-edge peaks of the XANES
lines (Fig.~\ref{fig:xanes}$(b)$,$(c)$). In both probed samples, there
are two defined peaks below the absorption edge, which we label A$_1$
and A$_2$, while the edge itself shows two shoulders, A$_3$ and
A$_{4}$. In Table~\ref{tab:xanes} the results of Gaussian fits
performed by using an arctan-function as baseline, are
reported. Similar findings were previously
interpreted\cite{Titov:2005_PRB,Antonov:2010_PRB} as indicative of the
Mn$^{3+}$ charge state. The peaks A$_1$ and A$_2$ correspond to the
transitions to Mn 3$d$-4$p$ hybrid states, while A$_3$ and A$_4$ end
in the GaN higher conduction bands at positions with a high density of
4$p$ states. Due to the tetrahedral environment, the Mn 3$d$-levels
split in two nearly degenerate $e$- and three nearly degenerate
$t_2$-levels for each spin-direction. The actual position of those
states with respect to the GaN band structure is still a matter of
debate, but from absorption\cite{Korotkov:2001_PBCM,Wolos:2004_PRB_a}
and photoluminescence\cite{Zenneck:2007_JAP} measurements it is known
that for the majority spin carriers in Mn$^{3+}$, the $e$-levels lie
around 1.4~eV below the $t_2$-levels of Mn incorporated
substitutionally in GaN, and the $t_2$-level, {\em i.e.,} the
Mn$^{3+}$/Mn$^{2+}$ state is located about 1.8~eV above the valence
band. An interpretation of simulations applied to x-ray absorption spectra
is given in Refs.~\onlinecite{Titov:2005_PRB} and \onlinecite{Titov:2006_thesis}, and
states that, due to crystal field effects, the 3$d$- and 4$p$-states
can hybridize, making transitions from the 1$s$-level to the
$t_2$-levels dipole allowed, while the interaction of the $e$-levels
with the 4$p$ orbitals is much weaker and cannot be seen in K-edge
XANES.

\begin{table}[tbp]
  \caption{Position $P$, integrated intensity $I$ and full width at half maximum $W$ of the Gaussians fitted to the peaks before and at the absorption edge. The background function is used to normalize the spectra.}
  \label{tab:xanes}
  \begin{center}
    \begin{tabular}{|l|ccc|ccc|}
      \hline
      \hline
      & \multicolumn{3}{c|}{100A} & \multicolumn{3}{c|}{490A}\\
      & $P$ (eV) & $I$ & $W$ (eV) & $P$ (eV) & $I$ & $W$ (eV)\\
      & $\pm$~0.2 & $\pm$~0.05 & $\pm$~0.1 & $\pm$~0.2 & $\pm$~0.05 & $\pm$~0.1\\
      \hline
      A$_1$ & 6538.9 & 0.18 & 1.5 & 6538.9 & 0.18 & 1.3 \\
      A$_2$ & 6540.8 & 0.32 & 1.6 & 6540.8 & 0.33 & 1.5 \\
      A$_3$ & 6545.8 & 0.70 & 3.6 & 6545.7 & 0.64 & 3.3 \\
      A$_4$ & 6548.7 & 0.28 & 2.1 & 6548.4 & 0.22 & 2.0 \\
      \hline
      \hline
    \end{tabular}
  \end{center}
\end{table}

In view of the above discussion we explain the physical mechanism
beyond the observed data considering possible $\textit{final}$ states
of the transitions from the $1s$ Mn shell. The final state
corresponding to the A$_1$ peak is Mn$^{2+}$, {\em i.e.} a $^6$A$_1$
state ($^6$S for the spherical symmetry), consisting of
$e^{2\uparrow}$ and $t_2^{3\uparrow}$ one electron levels. The A$_2$
peak can be interpreted as a crystal field multiplet derived from the
$^4G$ state consisting of
$e^{2\uparrow}t_2^{2\uparrow}t_2^{\downarrow}$, and lying about 2.5~eV
higher than the A$_1$ state. Apart from what reported in literature, a
reason why A$_1$ and A$_2$ are assigned to localized Mn-states is that
from the previous EXAFS analysis (Sec.~\ref{sec:exafs}) we obtain an
absorption edge value of 6543(1)~eV, between the energies of the A$_2$
and A$_3$ peaks, meaning that electrons excited to A$_1$ and A$_2$ can
not backscatter at the surrounding atoms, and they are thus
localized. This assignment gives a valuable information, namely, that
there is an empty state in the majority-spin $t_2$-level confirming
that most of the incorporated Mn-ions are really in the 3+ valence
state, in agreement with the conclusions of Refs.
\onlinecite{Sarigiannidou:2006_PRB} and \onlinecite{Titov:2005_PRB}. The model
explains also the presence of only one pre-edge peak in the case of
(Ga,Mn)As and $p$-(Zn,Mn)Te.\cite{Titov:2005_PRB} In those systems we
deal with Mn$^{2+}$ and delocalized holes, so that the final state of
the relevant transitions corresponds to the Mn $d^6$ level, involving
only one spin orientation. On the other hand, the XANES data do not
provide information on the radius of the hole localization in
(Ga,Mn)N, in other words, whether the Mn$^{3+}$ configuration
corresponds to the $d^4$ or rather to the $d^5$ + h situation, where
the relevant $t_2$ hole state is partly built from the neighboring
anion wave functions owing to a strong $p-d$ hybridization.

We also have simulated the Mn$_{Ga}$ K-edge absorption spectra in a
Ga$_{95}$Mn$_1$N$_{96}$ cluster (a $4a\times 4a\times 3c$ supercell,
corresponding to 1\% Mn concentration) focusing the attention on the
Mn electronic configuration: 3$d^4$ and 3$d^5$. The calculation is
conducted within the multiple-scattering approach implemented in {\sc
fdmnes}\cite{Joly:2001_PRB} using muffin-tin potentials, the
Hedin-Lunqvist approximation for their energy-dependent part, a
self-consistent potential calculation\cite{Joly:2009_JPC} for
enhancing the accuracy in the determination of the Fermi energy and
the in-plane polarization ($E \perp c$). Despite it is common practice
to report convoluted spectra to mimic the experimental resolution, we
find out that this procedure can arbitrary change the layout of the
pre-edge peaks and for this reason it is preferred to show
non-convoluted data [Fig.~\ref{fig:xanes}$(b)$)]. Regarding the fine
structure of the simulated spectra, we have a good agreement with
experimental data, confirming the Mn$_{\rm Ga}$ incorporation as found
by the EXAFS analysis (Sec.~\ref{sec:exafs}). On the other hand, the
simulated pre-edge features need a further investigation: the
experimental intensity of A$_1$ and A$_2$ and the position of A$_3$
are not properly reproduced. This could be due to some neglected
effects in the employed formalism, as explained in
Ref.~\onlinecite{Titov:2005_PRB}, where the two peak structure was
reproduced theoretically within a more elaborated model.

\section{Magnetic properties}
\label{sec:magnetic_properties}
\subsection{SQUID results}\label{sec:squid}

\begin{figure}[b]
  \centering
  \includegraphics[width=8.5cm]{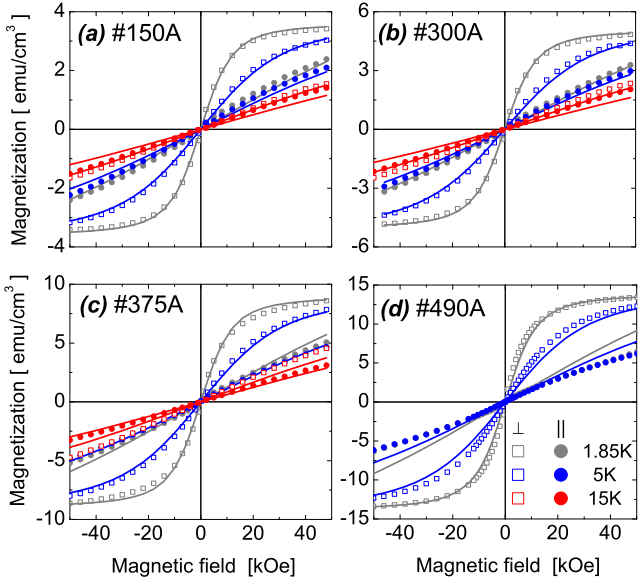}
  \caption{(Color online) Magnetization measurements at 1.85, 5, and
    15~K of Ga$_{1-x}$Mn$_x$N as a function of the magnetic field
    applied parallel (closed circles) and perpendicular (open squares)
    to the GaN wurtzite $c$-axis. The solid lines show the
    magnetization curves calculated according to the group theoretical
    model for non-interacting Mn$^{3+}$ ions in wz-GaN.}
  \label{fig:M_H_diffT}
\end{figure}

\begin{figure}[htbp]
  \centering
  \includegraphics[width=7.5cm]{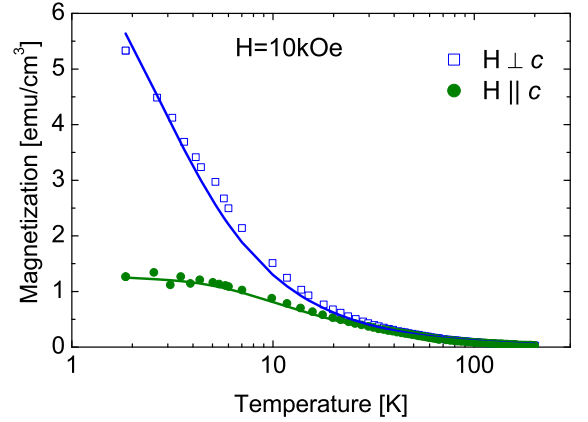}
  \caption{(Color online) Temperature dependence of the magnetization
    $M$ for sample 375A (points) at $H=10$~kOe. The solid lines
    represent the magnetization calculated within the group theoretical
    model of non-interacting Mn$^{3+}$ ions in wz-GaN.}
  \label{fig:M_T}
\end{figure}

\begin{figure}[tbp]
  \centering
  \includegraphics[width=7.5cm]{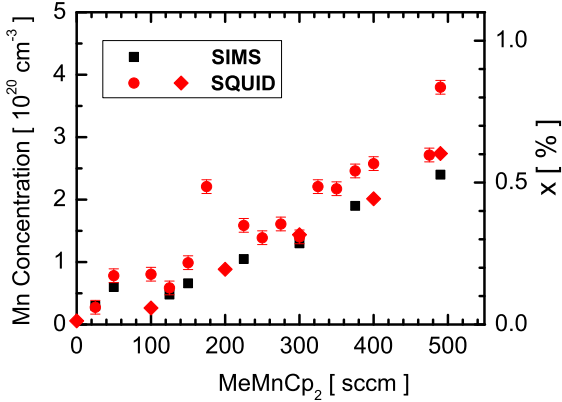}
  \caption{(Color online) Mn concentration $n_{\mathrm{Mn}}$ obtained
    from magnetization measurements (circles - series A, diamonds -
    series B) and SIMS (squares - series A) as a function of the Mn
    precursor flow rate.}
  \label{fig:Mn_Concentration}
\end{figure}

We investigate both the temperature dependence of the magnetization
$M$ at a constant field $M(T)$ and the sample response to the
variation of the external field at a constant temperature $M(H)$. The
same experimental routine is repeated for both in-plane and
out-of-plane configurations, that is with magnetic field applied
perpendicular and parallel to the hexagonal $c$-axis, respectively. In
Fig.~\ref{fig:M_H_diffT} representative low temperature $M(H)$ data
for both orientations are reported. We note that these curves exhibit
a paramagnetic behavior with a pronounced anisotropy with respect to
the $c$-axis of the crystal.  This indicates a nonspherical Mn ion
configuration, expected for a $L\neq0$ state. At the same time we
report an absence of any ferromagnetic-like features, that---on the
other hand---are typical for (Ga,Fe)N layers\cite{Bonanni:2007_PRB} at
these concentrations of the magnetic ions, supporting the absence of
crystallographic phase separation in our layers, as suggested by the
SXRD and HRTEM studies. The same finding additionally indicates that
both chemical phase separation (spinodal decomposition) and
medium-to-long range ferromagnetic spin-spin coupling are also absent
in this dilute layers. The latter allows us to treat the Mn ions as
completely non-interacting, at least in the first approximation. The
solid lines in Figs.~\ref{fig:M_H_diffT} and \ref{fig:M_T} represent
fits to our experiential data on the paramagnetic response of
non-interacting Mn$^{3+}$ ions ($L=2$, $S=2$) with the trigonal
crystal field of the wurzite GaN structure and the Jahn-Teller
distortion taken into account (details in Sec.~\ref{sec:CF}). The
overall match validates our approach, which, in turn, is consistent
with previous findings\cite{Graf:2002_APL,Graf:2003_PSSB} that without
an intentional codoping, or when the stoichiometry of GaN:Mn is
maintained, Mn is occupying only the neutral Mn$^{3+}$ acceptor
state. Interestingly, all theoretical lines in
Figs.~\ref{fig:M_H_diffT} and \ref{fig:M_T} are calculated employing
only one set of crystal field parameters (as listed in
Table~\ref{tab:Parameters_CF}) having the Mn$^{3+}$ concentration
$n_{\mathrm{Mn^{3+}}}$ as the only adjustable parameter for each
individual layer. In Fig.~\ref{fig:Mn_Concentration} the
$n_{\mathrm{Mn^{3+}}}$ values as a function of the manganese precursor
flow rate are given together with the total Mn content
$x_{\mathrm{Mn}}$ as determined by SIMS.

\begin{figure}[htbp]
  \centering
  \includegraphics[width=7.5cm]{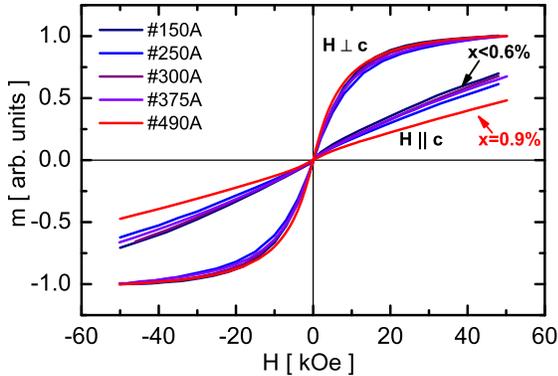}
  \caption{(Color online) Magnetization curves at T=1.85K of
    five Ga$_{1-x}$Mn$_x$N samples normalized with respect to their in
    plane magnetization at $H=50$~kOe.}
  \label{fig:M_H_normalized}
\end{figure}

However, there are hints that the interaction between Mn spins may
play a role for $x \gtrsim 0.6$\%.  In Fig.~\ref{fig:M_H_normalized}
the $M(H)$ normalized at high field ($H=50$~kOe) to their in-plane
values, are reported. The fact that the shape of their magnetization
curves is independent of $x$ for $x \lesssim 0.6$\% means that the
interactions between Mn ions are unimportant for these dilutions. On
the other hand, the $M(H)$ for a layer with $x = 0.9$\% (490A)
secedes markedly from the curves for samples with $x \lesssim 0.6$\%,
indicating that supposedly ferromagnetic Mn-Mn coupling starts to
emerge with increasing relative number of Mn nearest neighbors in the
layers.  Nevertheless, due to the generally low Mn concentration in the
considered samples, no conclusive statement about the strength of the
magnetic couplings can be drawn from our magnetization data.
Interestingly, depending on the very nature of the Mn centers both
ferromagnetic and/or antiferromagnetic $d$-$d$ interactions can emerge
in (Ga,Mn)N. The presence of Mn$^{2+}$ ions essentially leads to
antiferromagnetic superexchange, as in II-Mn-VI DMS, where
independently of the electrical doping, the position of the Mn
$d$-band guarantees its 3$d^5$ configuration. Significantly, the same
antiferromagnetic $d$-$d$ ordering and paramagnetic behavior typical
for $S=5/2$ of Mn$^{2+}$ was reported in $n$-type bulk (Ga,Mn)N samples
containing as much as $9\%$ of Mn.\cite{Zajac:2001_APL} On the other
hand, calculations for Mn$^{3+}$ within the DFT point to ferromagnetic
coupling\cite{Boguslawski:2005_PRB,Cui:2007_PRB} and, experimentally,
a Curie temperature $T_{\mathrm{C}}\simeq 8~K$ was observed in
single-phase Ga$_{1-x}$Mn$_x$N with $x \simeq 6\%$ and the majority of
Mn atoms in the Mn$^{3+}$ charge
state.\cite{Sarigiannidou:2006_PRB,Marcet:2006_PSSC} Our experimental
data seems to support these findings and to extend their validity
towards the very diluted limit. Finally, we remark that the
carrier-mediated ferromagnetism can be excluded at this stage due to
the insulating character of the samples, confirmed by room
temperature four probe resistance measurements and consistent with the
mid-gap location of the Mn acceptor level.

The observations presented here point to an uniqueness of Mn in
GaN. The fact that Ga$_{1-x}$Mn$_x$N with $x \lesssim 1$\% is
paramagnetic without even nanometer-scale ordering should be
contrasted with GaN doped with other TM ions. Depending on the growth
conditions, the TM solubility limit is rather low and typically,
except for Mn, it is difficult to introduce more than $1\%$ of magnetic
impurities into randomly distributed substitutional sites. For
example, the solubility limit of Fe in GaN has been shown to be
$x\approx0.4\%$ at optimized growth conditions (see
Ref.~\onlinecite{Bonanni:2007_PRB}), but signatures of a nanoscale
ferromagnetic coupling are observed basically for any
dilution.\cite{Bonanni:2007_PRB} The relatively large solubility limit
of Mn in GaN, in turn, has a remarkable significance in the search for
long-range coupling mediated by itinerant carriers.\cite{Dietl:2000_S,
  Dietl:2001_PRB} Not only it lets foresee a high concentration of
substitutional Mn---important for the long-range ordering---but it can
ensure that the effects brought about by carriers are not masked by
signals from nanocrystals with different phases.

\subsection{Magnetism of Mn$^{3+}$ ions - theory}
\label{sec:CF}

The Mn concentration in our samples is then $x\lesssim 1\%$ as
evaluated by means of various characterization techniques
(subsection~\ref{sec:x-Mn}), implying that most of the Mn ions ($\geq
90\%$) have no nearest magnetic neighbors. Therefore, the model that
considers the Mn ions as single, noninteracting magnetic centers
should provide a reasonable picture. To describe the Mn$^{3+}$ ion we
follow the group theoretical model developed for Cr$^{2+}$ ion by
Vallin\cite{Vallin:1970_PRB, Vallin:1974_PRB} and then successfully
used for a Mn-doped hexagonal GaN
semiconductor.\cite{Gosk:2005_PRB,Wolos:2004_PRB_b} It should be
pointed out that symmetry considerations cannot discriminate between
the $d^5$+hole and $d^4$ many-electron configurations of the Mn ions,
therefore the presented model should be applicable to both
configurations. Through this section, the capital letters $T_i$
($i=1,2$), $E$ denote the irreducible representations of the point
group for the multielectron configurations in contrast to the single
electron states indicated by the small letters $d$, $e$, $t_2$.

We consider a Mn ion that in its free state is in the electronic
configuration $d^5s^2$ of the outer shells. When substituting for the
group III ($s^2p$) cation site, Mn gives three of its electrons to the
crystal bond and assumes the Mn$^{3+}$ configuration. In a tetrahedral
crystal field, the relevant levels are five-fold degenerate with
respect to the projection of the orbital momentum and are splet by
this field and by hybridization with the host orbitals into two
sublevels $e$ and $t_2$ with different energies. In the tetrahedral
case the $e$ states lie lower than the $t_2$ states. This fact can be
understood by analyzing the electron density distribution of the
$t_2^{xy}$, $t_2^{yz}$, $t_2^{zx}$ and $e^{x^2-y^2}$, $e^{z^2}$
levels. The density of the $t_2$ state extends along the direction
toward the N ligand anions, while the $e$ orbital has a larger
amplitude in the direction maximizing the distance to the N ion and
due to the negative charge of the N anions, the $t_2$ energy
increases. However, the relevant, {\em i.e.}, the uppermost $t_2$
state may actually originate from orbitals of neighboring anions, pull
out from the valence band by the $p$-$d$
hybridization.\cite{Dietl:2008_PRB} If the system has several
localized electrons, they successfully occupy the levels from the
bottom, according to Hund's first rule, and keep their spins
parallel. By considering the full orbital and spin moments, the
Mn$^{3+}$ center can be described through the following set of quantum
numbers ($Lm_LSm_S$) with $L=2$ and $S=2$. However, we underline that
this procedure can be used only if the intra-atomic exchange
$\Delta_{ex}$ interaction is larger than the splitting between the $e$
and $t_2$ states $\Delta_{CF}=E_{t_2}-E_e$ ($\Delta_{ex} >
\Delta_{CF}$). After this, the effect of the host crystal is taken
into account as a perturbation like in the single electron
problem. One forms first $2L+1$ wave functions for the $n$-electron
system determined by the Hund's rule, calculates the matrix elements
for these states and determines the energy level structure. In this
way, the impurity ions states are found and classified according to
the irreducible representations of the crystal point group and
characterized by the set ($\Gamma MSm_S$) of quantum numbers, with $M$
the number of the line of an irreducible representation $\Gamma = A_1,
A_2, E, T_1, T_2$ of the corresponding point group. In the case of a
Mn$^{3+}$ ($L=2, S=2$) ion in a tetrahedral environment the ground
state corresponds to the $^5T_2(e^2t_2^2)$ configuration with two
electrons in the $e$ and two electrons in the $t_2$ level. The ground
state is three-fold degenerate, since there are three possibilities to
choose two orbitals from three $t_2$ orbitals. The first excited state
for the Mn$^{3+}$ ion is $^5E(e^1t_2^3)$ (see
Ref.~\onlinecite{Kikoin:2004_B}).

\begingroup
\squeezetable

\begin{table}[tbp]
  \centering \caption{Parameters of the group theoretical model used to calculate the magnetization of Ga$_{1-x}$Mn$_x$N. All values are in meV.}
  \begin{ruledtabular}
    \begin{tabular}{ ccccccc}

      $B_4$ &  $B_2^0$   & $B_4^0$  &  $\tilde{B}_2^0$ & $\tilde{B}_4^0$ & $\lambda_{TT}$ & $\lambda_{TE}$ \\

      \hline

      \\

      11.44&4.2&-0.56&-5.1&-1.02&5.0&10.0\\

    \end{tabular}
  \end{ruledtabular}
  \label{tab:Parameters_CF}
\end{table}
\endgroup

The energy structure of a single ion in Mn$^{3+}$ charge state can be
described by the Hamiltonian

\begin{eqnarray}
\label{eq:Hcf}
H=H_{\mathrm{CF}}+H_{\mathrm{JT}}+H_{\mathrm{TR}}+H_{\mathrm{SO}}+H_{\mathrm{B}},
\end{eqnarray}

where $H_{\mathrm{CF}}=-2/3B_4(\hat{O}_4^0-20\sqrt{2}\hat{O}_4^3)$
gives the effect of a host having tetrahedral $T_{d}$ symmetry,
$H_{\mathrm{JT}}=\tilde{B}_2^0\hat{\Theta}_4^0+\tilde{B}_4^0\hat{\Theta}_4^2$
is the static Jahn-Teller distortion of the tetragonal symmetry,
$H_{\mathrm{TR}}=B_2^0\hat{O}_4^0+B_4^0\hat{O}_4^2$ represents the
trigonal distortion along the GaN hexagonal $c$-axis, that lowers the
symmetry to $C_{3V}$, $H_{\mathrm{SO}}=\lambda\hat{L}\hat{S}$
corresponds to the spin-orbit interaction and
$H_B=\mu_B(\hat{L}+2\hat{S})\textbf{B}$ is the Zeeman term describing
the effect of an external magnetic field. Here $\hat{\Theta}$,
$\hat{O}$ are Stevens equivalent operators for a tetragonal distortion
along one of the cubic axes $[100]$ and trigonal axis $[111]\|c$ (in a
hexagonal lattice) and $B_q^p$, $\tilde{B}_q^p$, $\lambda_{TT}$, and
$\lambda_{TE}$ are parameters of the group theoretical model. As
starting values we have used the parameters reported for Mn$^{3+}$ in
GaN:Mn,Mg \cite{Wolos:2004_PRB_b} which describe well the
magneto-optical data on the intra-center absorption related to the
neutral Mn acceptor in GaN. Remarkably, only a noticeable
modification (about 10\%) of $\lambda_{TT}$ and $B_2^0$ has been
necessary in order to reproduce our magnetic data (the remaining
parameters are within 3\% of their previously determined values.)
Actually, the model with the parameter values collected in Table III
describes both the magnetization $M(H)$ and its crystalline anisotropy
(Figs.~\ref{fig:M_H_diffT} and \ref{fig:M_T}) as well as the position and the field-induced splitting of optical lines.\cite{Wolos:2004_PRB_b}

The ground state of the Mn$^{3+}$ ion is an orbital and spin quintet
$^5D$ with $L=2$ and $S=2$. The term $H_{\mathrm{CF}}$ splits the
$^5D$ ground state into two terms of symmetry $^5E$ and $^5T_2$
(ground term). The $^5E-^5T_2$ splitting is $\Delta_{CF}=120B_4$. The
nonspherical Mn$^{3+}$ ion undergoes further Jahn-Teller distortion,
that lowers the local symmetry and splits the ground term $^5T_2$ into
an orbital singlet $^5B$ and an higher located orbital doublet
$^5E$. The trigonal field splits the $^5E$ term into two orbital
singlets and slightly decreases the energy of the $^5B$ orbital
singlet. The spin-orbital term yields further splitting of the spin
orbitals. Finally, an external magnetic field lifts all of the
remaining degeneracies.

For the crystal under consideration, there are three Jahn-Teller
directions: $[100]$, $[010]$ and $[001]$ (center A, B, C
respectively).\cite{Gosk:2005_PRB, Wolos:2004_PRB_b} It should be
pointed out that the magnetic anisotropy of the Mn$^{3+}$ system
originates from different distributions of nonequivalent Jahn-Teller
centers in the two orientations of the magnetic field and the
hexagonal axial field $H_{TR}$ along the $c$-axis. This picture of Mn
in GaN emphasizing the importance of the Jahn-Teller effect, which  lowers the local  symmetry and splits the ground term $^5T_2$ into an orbital singlet and a doublet, is in agreement with a recent $\textit{ab initio}$ study employing a hybrid exchange potential.\cite{Stroppa:2009_PRB}

The energy level scheme of the Mn$^{3+}$ ion is calculated through a
numerical diagonalization of the full $25\times 25$ Hamiltonian
(\ref{eq:Hcf}) matrix. The average magnetic moment of the Mn ion
$\textbf{m}=\textbf{L}+2\textbf{S}$ (in units of $\mu_B$) can be
obtained according to the formula:

\begin{equation}
\label{eq:M_cf}
<\textbf{m}>=Z^{-1}(Z_A<\textbf{m}>^A+Z_B<\textbf{m}>^B+Z_C<\textbf{m}>^C),
\end{equation}

with $Z_i$ ($i=A, B$ or $C$) being the partition function of the i-th
center, $Z=Z_A+Z_B+Z_C$ and

\begin{equation}
\label{eq:M_cf_Center}
<\textbf{m}>^i=\frac{\sum_{j=1}^N<\varphi_{j}|\hat{L}+2\hat{S}|\varphi_{j}>\mathrm{exp}(-E_j^i/k_BT)}{\sum_{j=1}^N\mathrm{exp}(-E_j^i/k_BT)},
\end{equation}

where $E_j^i$ and $\varphi_{j}$ are the j-th energy level and the
eigenstate of the Mn$^{3+}$ ion $i$-th center, respectively. As
already mentioned, the Mn concentration in our samples is relatively
small $x\lesssim 1\%$. Therefore, the model assuming a system of
single Mn ions provides a reasonable description of the magnetic
behavior. The macroscopic magnetization \textbf{M}, shown in
Figs.~\ref{fig:M_H_diffT} and \ref{fig:M_T}, can then be expressed in
the form

\begin{equation}
\label{eq:m_macro}
\textbf{M}=\mu_B<\textbf{m}>n_{Mn},
\end{equation}

where $n_{\mathrm{Mn}}=N_{\mathrm{Mn}}/V$ is the Mn concentration and
$N_\mathrm{Mn}$ the total number of Mn ions in a volume $V$.
\subsection{Search for hole-mediated ferromagnetism}

As already mentioned, according to the theoretical predictions within
the $p-d$ Zener model,\cite{Dietl:2000_S, Dietl:2001_PRB} RT
ferromagnetism is expected in single-phase (Ga,Mn)N and related
compounds, provided that a sufficiently high concentration of both
substitutional magnetic impurities (near 5$\%$ or above) and
valence-band holes will be realized. The latter condition is a more
severe one, as the high binding energy of Mn acceptors in the strong
coupling limit leads to hole localization.

\begin{figure}[b]
  \centering
  \includegraphics[width=7.5cm]{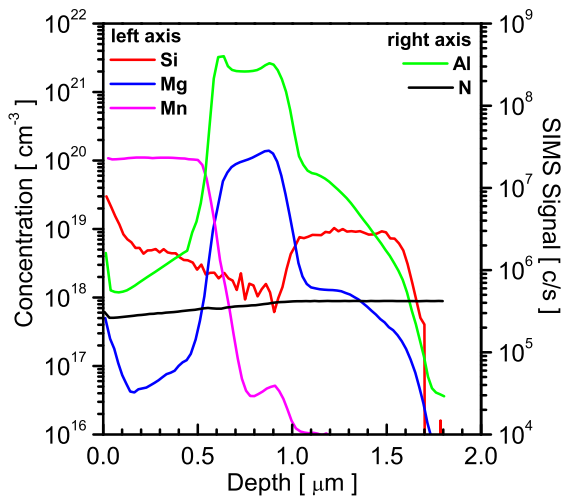}
  \caption{(Color online) SIMS depth profiles of our (Ga,Mn)N/(Al,Ga)N:Mg/GaN:Si ($i$-$p$-$n$)
    structure.}
  \label{fig:nepal_sims}
\end{figure}

Surprisingly, RT ferromagnetism in $p$-type Ga$_{1-x}$Mn$_x$N with a Mn
content as low as $x \approx 0.25\%$ was recently
reported.\cite{Nepal:2009_APL} The investigated modulation-doped
structure consisted of a (Ga,Mn)N/(Al,Ga)N:Mg/GaN:Si ($i$-$p$-$n$)
multilayer and a correlation between the ferromagnetism of the
(Ga,Mn)N film at 300~K and the concentration of holes accumulated at
the (Ga,Mn)N/(Al,Ga)N:Mg interface was shown. The interfacial hole
density was controlled by an external gate voltage applied across the
$p$-$n$ junction of the structure, and a suppression of the FM
features---already existing without the gate bias---took place for a
moderate gate voltage applied. Apart from a high value of
$T_{\mathrm{C}}$, a puzzling aspect of the experimental results is the
large magnitude of the spontaneous magnetization,
$75$~$\mu$emu/cm$^2$.\cite{Nepal:2009_APL} Since the holes are
expected to be accumulated in a region with a thickness of the order
of 1~nm, the reported magnetic moment is about two orders of magnitude
larger than the one expected for ferromagnetism originating from an
interfacial region in (Ga,Mn)N with $x = 0.25$\%.

\begin{figure}[htbp]
  \centering
  \includegraphics[width=7.5cm]{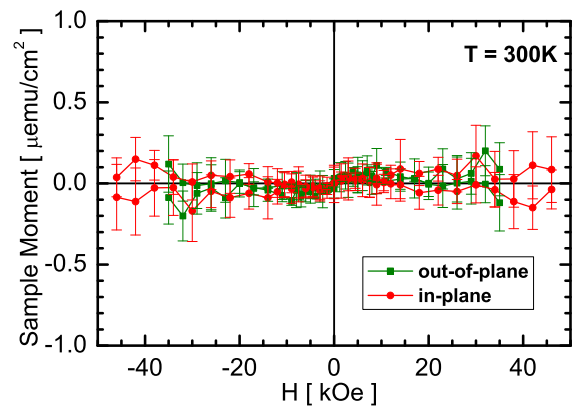}
  \caption{(Color online) Room temperature magnetic signal from the
    (Ga,Mn)N/(Al,Ga)N:Mg/GaN:Si structure. For completeness, results
    of both in-plane and out-of-plane orientations are
    shown. Diamagnetic and paramagnetic contributions have been
    compensated.}
  \label{fig:nepal_squid}
\end{figure}

Nevertheless, we have decided to check the viability of this approach
that not only seemed to result in high temperature FM in GaN:TM, but
also allowed the all-electrical control of FM. Thus, we have combined
the $p$-type doping procedures we previously
optimized\cite{Simbrunner:2007_APL} with the growth of the dilute
(Ga,Mn)N presented in this work to carefully reproduce the
corresponding structure.\cite{Nepal:2009_APL} The desired architecture
of the investigated sample is confirmed by SIMS profiling (see
Fig.~\ref{fig:nepal_sims}) indicating the formation of well defined
(Ga,Mn)N/(Al,Ga)N:Mg and (Al,Ga)N:Mg/GaN:Si interfaces. However, as
shown in Fig.~\ref{fig:nepal_squid}, no clear evidence of a
ferromagnetic-like response is seen within our present experimental
resolution of $\approx 0.3$~$ \mu\mbox{emu/cm}^{2}$. To strengthen
the point, we note here that the maximum error bar of our results
($\approx 0.7$~$ \mu\mbox{emu/cm}^{2}$) corresponds to about 1/100 of
the saturation magnetization reported in the assessed experiment.
While the absence of a ferromagnetic response at the level of our
sensitivity is to be expected, the presence of a large ferromagnetic
signal found in Ref.~\onlinecite{Nepal:2009_APL} in a nominally
identical structure is surprising. Without a careful structural
characterization of the sample studied in
Ref.~\onlinecite{Nepal:2009_APL} by methods similar to those we have employed
in the case of our layers, the origin of differences in magnetic
properties between the two structures remains unclear.

\section{Summary}
\label{sec:Summary}

In this paper we have investigated Ga$_{1-x}$Mn$_x$N films grown by
MOVPE with manganese concentration $x\lesssim 1\%$. A set of
experimental methods, including SXRD, HRTEM, and EXAFS, has been
employed to determine the structural properties of the studied
material. These measurements reveal the absence of crystallographic
phase separation and a Ga-substitutional position of Mn in GaN. The
findings demonstrate that the solubility of Mn in GaN is much greater
than the one of Cr (Ref.~\onlinecite{Cho:2009_JCG}) and Fe
(Ref.~\onlinecite{Bonanni:2008_PRL}) in GaN grown under the same
conditions. Nevertheless, for the attained Mn concentrations and owing
to the absence of band carriers, the Mn spins remain
uncoupled. Accordingly, pertinent magnetic properties as a function of
temperature, magnetic field and its orientation with respect to the
$c$-axis of the wurtzite structure can be adequately described by the
paramagnetic theory of an ensemble of non-interacting Mn ions in the
relevant crystal field. Our SQUID and XANES results point to the 3+
configuration of Mn in GaN. However, the collected information can not
tell between $d^4$ and $d^5 + h$ models of the Mn$^{3+}$ state, that
is on the degree of hole localization on the Mn ions. A negligible
contribution of Mn in the 2+ charge state indicates a low
concentration of residual donors in the investigated films. Our
studies on modulation doped $p$-type Ga$_{1-x}$Mn$_{x}$N/(Ga,Al)N:Mg
heterostructures do not reproduce the high temperature robust
ferromagnetism reported recently for this system.\cite{Nepal:2009_APL}

\section*{Acknowledgements}

The work was supported by the FunDMS Advanced Grant of the European
Research Council within the "Ideas" 7th Framework Programme of the EC,
and by the Austrian Fonds zur {F\"{o}rderung} der wissenschaftlichen
Forschung (P18942, P20065 and N107-NAN). We also acknowledge H.~Ohno
and F.~Matsukura for valuable discussions, G. Bauer and R.T. Lechner
for their contribution to the XRD measurements as well as the support
of the staff at the Rossendorf Beamline (BM20) and at the Italian
Collaborating Research Group at the European Synchrotron Radiation
Facility in Grenoble.

\bibliographystyle{apsrev}

\bibliography{Baza_nowa}

\end{document}